\documentclass[peerreview,onecolumn,amsmath,amssymb]{IEEEtran}
\usepackage{rotating}
\usepackage{graphicx,float}
\usepackage{color}
\usepackage{amsmath}
\usepackage{amssymb}
\usepackage[abs]{overpic}     % latex on top of a picture

\begin{document}

\title{Ultrasound imaging with three dimensional full-wave nonlinear acoustic simulations.\\ Part 2: sources of image degradation in intercostal imaging.}

\author{Gianmarco~Pinton$^1$\\$^1$Joint Dept. of Biomedical Engineering, University of North Carolina at Chapel Hill and North Carolina State University\\}
\thanks{Further author information: \\Send correspondence to G. Pinton: E-mail: gia@email.unc.edu}
 
\markboth{}{Pinton: }
\maketitle

\newcommand{\invivo}{\emph{in vivo}} % use \invivo\ to add a space after vivo
\newcommand{\exvivo}{\emph{ex vivo}}
\newcommand{\insilico}{\emph{in silico}}
\newcommand{\insitu}{\emph{in situ}}

%%%%%%%%%%%%%%%%%%%%%%%%%%%%%%%%%%%%%%%%%%%%%%%%%%%%%%%%%%%%%
\begin{abstract}
Full-wave simulations are applied to an intercostal imaging scenario to determine the sources of fundamental and harmonic image degradation with respect to aberration and reverberation. These simulations are based on Part I of this two part paper, which established the full-wave simulation methods to generate realistic ultrasound images based directly on the first principles of wave propagation in the human body. The ultasound images are generated based on the first principles of propagation and reflection. By relying on first priniples of the three dimensional wave propagation physics the interplay between distributed aberration and reverberation clutter can be fully appreciated. Three imaging scenarios that would not be realizable {\it in vivo} are investigated {\it in silico}. First, the ribs were completely removed and replaced with fat. Then, the ribs were maintained in their anatomically correct configuration to yield a reference image. Finally the ribs were placed closer together in elevation. The propagation based B-mode images show that of these three scenarios the second, anatomically correct configuration, has the best contrast-to-noise ratio. This is due to two competing effects. First it is shown that the ribs effectively apodize the fundamental and harmonic beams by 3-5 dB. This effect alone would predict an improvement in image quality. However, the B-mode image quality, measured by the contrast-to-noise ratio degrades by 8\%.  To fully explain these changes, it is shown that a second effect, multiple reverberation, must be taken into account. A point spread function analysis shows that when the ribs are placed closer together they also generate significantly more reverberation clutter (by 2.4 to 2.9 dB), which degrades the image quality even though the beamplot has lower sidelobes. In this intercostal imaging scenario the effects of the ribs on beam shape and reverberation are therefore in competition in terms of image quality and there is an optimal acoustic window that balances them out. This simulation tool and image quality analysis could be applied to other imaging configurations and in other areas of the body.
\end{abstract}

\pagebreak

\section{Introduction}

Intercostal imaging is challenging due to the restrictive anatomy
which limits the imaging window and due to the aberration and
reverberation introduced by the body wall and ribs. The primary
applications of intercostal imaging are to cardiac and liver scanning.
Two major developments have improved intercostal imaging, high channel
count 2D or matrix arrays and harmonic
imaging~\cite{szabo2013ultrasound}. Matrix arrays can dynamically
focus in the elevation and lateral planes, which improves the
resolution. Harmonic imaging improves image quality by reducing the
influence of multipath reverberation from the body
wall~\cite{pinton2011sources}.

In a companion paper we established the methodology to link three
dimensional acoustical maps of the human body to image quality
metrics. We showed that full-wave simulations can accurately describe
the ultrasound imaging physics based directly on the first principles
of wave propagation in heterogeneous media. In this paper we will look
at intercostal imaging and how this simulation tool, called
``Fullwave'' ~\cite{pinton2009heterogeneous} can determine the sources
of image degradation in an intercostal imaging scenario that would not
be realizable {\it in vivo}.

Several groups have developed methods to determine the influence of
ribs on ultrasound propagation for imaging and therapy. In an {\it ex
  vivo} experiment Hinkelman {\it et
  al.}~\cite{hinkelman1997measurements} determined the deviation of
the arrival time across a mechanically scanned 2-D aperture for
ultrasound propagating through a human chest wall. These
root-mean-square estimates were, on average, 21.3 ns.

For a transcostal high-intensity, focused ultrasound (HIFU) study
Aubry {\it et al.}~\cite{aubry2008transcostal} used a time reversal
mirror~\cite{fink1992time} to focus behind porcine ribs. This resulted
in a 5 dB reduction of the sidelobes and a six-fold increase in energy
delivered at the focus. Simulations performed by Bobkova {\it et
  al.}~\cite{bobkova2010focusing} compared time-reversal and geometric
approaches showing that both approaches improved the focal energy
deposition by approximately a factor of two.  An adaptive focusing
technique implemented on a combined imaging and therapy array by
Ballard {\it et al.}~\cite{ballard2010adaptive} improved the efficacy
of transcostal HIFU by maximizing the target echogenicity. This
generated increased the temperature at the target by a factor of 1.7
to 6.7 times {\it in vivo}. Focal splitting in transcostal HIFU was
studied in simulations and experimentally by Khoklova {\it et
  al}~\cite{khokhlova2010focus}. It was shown that the ribs can cause
the formation secondary foci due to diffraction. Gelat {\it et
  al.}~\cite{gelat2011modelling} also observed secondary foci using a
boundary-element approach.

For imaging applications, Li {\it et al.}~\cite{li1993blocked}
developed a least-squares method to estimate and suppress unwanted
contributions from elements of the imaging array that were blocked by
the ribs. This method was shown to improve the image quality of point
target phantoms. Jakovljevic {\it et
  al.}~\cite{jakovljevic2017blocked-a}, used the Fullwave simulation
tool and a clinical ultrasound scanner to characterize the signals
from a fully sampled matrix array. It was shown that blocked elements
had a lower amplitude and lower nearest-neighbor
correlation. Furthermore adding blocked elements to the beamformer
reduced vessel contrast by 19\% and vessel contrast to noise ratio by
10\%. Subsequently a blocked element compensation method was proposed
and shown to reduce reverberation clutter by 5
dB~\cite{jakovljevic2017blocked-b}. Using the two dimensional Fullwave
simulation tool in the human abdomen we have previously shown that
reverberation clutter is a principal source of image degradation at
the fundamental frequency~\cite{pinton2011sources}. These methods and
this approach is extended to three dimensions here.

%Blocked elements, coherence, ballard, etc 

%Jakovlevic

%SImulation studies: khoklova
%Gelat

The objective of this paper is to investigate the sources of image
degradation in intercostal imaging. The numerical and simulation
methods, described in detail in the companion paper, established
highly realistic acoustical maps of the human body based on the female
cadaver cryosections from the Visible
Human~\cite{spitzer1996visible}. These three dimensional simulations
provide a platform for understanding which mechanisms are
responsible for observed changes in image quality metrics.

We investigate imaging scenarios that are not realizable {\it in
  vivo}. The ribs are moved closer together, or removed altogether,
while maintaining all other imaging parameters, such as scatterer
resolution, body wall composition, and tissue distribution
unchanged. The effect of the ribs on image quality can therefore be
separated from potential confounding factors. Furthermore, since the
simulations are based on the full-wave propagation physics, the
acoustical field inside the body is known throughout the imaging
volume, which would not be possible {\it in vivo}.

By using estimates of the acoustical field in the human abdomen in
conjunction with point spread function simulations, lesion
detectability is linked to sources of image degradation as measured by
the aberration, reverberation clutter, and beamplots. Contrast to
noise estimates of the anechoic lesion are generated from B-mode
images based on transmit-receive simulations.

%designing better ultrasound
%imaging

%description of the human

%illustrate certain effects: reverberation stronger than aberration. That in some cases the ribs improve image quality and how.

%provides an ideal imaging scenario, ribs, aberration, multiple reverberation, challenging, 3D because of elevation, also forward looking because 3d, high channel count scanners, cardiac applications.

%Background and
%motivation Talk about why 3d is necessary, rib example. Higher channel
%count scanners. Simulations can yield controlled conditions inside the
%human body that would not be realizeable experimentally. Can also
%measure the field inside the body which yields information that you
%wouldn't be able to obtain otherwise.

\section{Methods}
A full description of the simulation methods and beamforming approach
is described the companion paper. Here the focus is on the application
of this tool for the simulation of a trancostal imaging scenario.

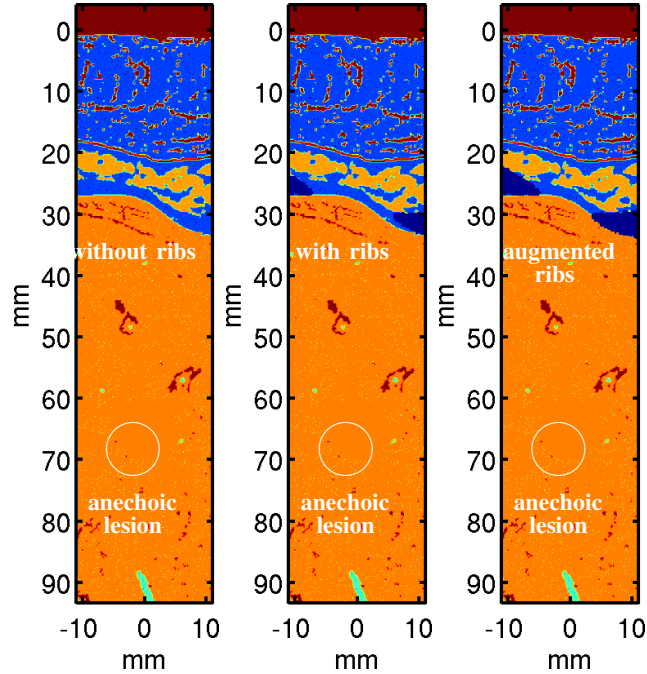
\begin{figure}[h]
\centerline{
 \begin{overpic}[width=0.15\linewidth]{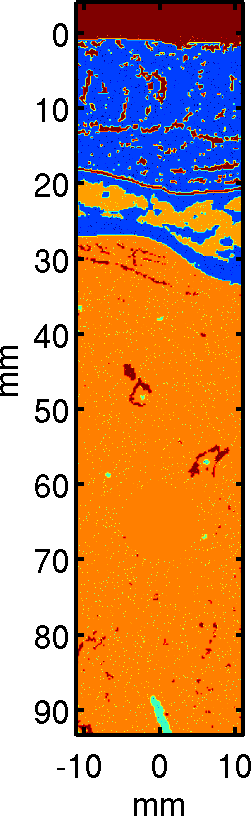}
     \put(45,83){\bf\textcolor{white}{\circle{20}}}
      \put(28,60){\small\bf\textcolor{white}{ anechoic }}
      \put(34,51){\small\bf\textcolor{white}{ lesion }}
      \put(21,155){\small\bf\textcolor{white}{ without ribs}}
 \end{overpic}
 \begin{overpic}[width=0.15\linewidth]{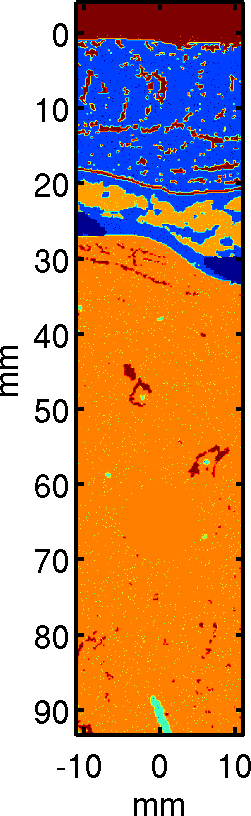}
     \put(45,83){\bf\textcolor{white}{\circle{20}}}
      \put(28,60){\small\bf\textcolor{white}{ anechoic }}
      \put(34,51){\small\bf\textcolor{white}{ lesion }}
\put(26,155){\small\bf\textcolor{white}{ with ribs}}
\end{overpic}
 \begin{overpic}[width=0.15\linewidth]{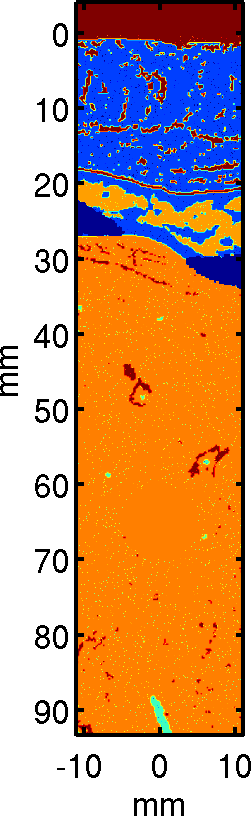}
     \put(45,83){\bf\textcolor{white}{\circle{20}}}
      \put(28,60){\small\bf\textcolor{white}{ anechoic }}
      \put(34,51){\small\bf\textcolor{white}{ lesion }}
\put(24,155){\small\bf\textcolor{white}{ augmented }}
\put(36,146){\small\bf\textcolor{white}{ribs}}
\end{overpic}}
 \caption{A slice along the elevation plane of the three dimensional
   speed of sound maps images of the human body illustrating three
   different configurations for the ribs. From left to right, no ribs,
   with anatomical rib positions, and with artificially augmented
   ribs.}
  \label{fig:ribs}
\end{figure}

Three rib configurations were modeled. Fig.~\ref{fig:ribs} illustrates
the three speed of sound maps in the elevation dimension, i.e. the
direction of the long axis of the transducer is into the page, along
the direction of the ribs. First, the ribs were maintained in their
anatomically correct configuration to yield a reference imaging
scenario (Fig.~\ref{fig:ribs}, middle). Second, the ribs were
completely removed and replaced with fat, which is the equivalent of
imaging through an area of the abdominal body without ribs
(Fig.~\ref{fig:ribs}, left).  This hypothetically generates a best
case scenario for intercostal imaging while retaining the degrading
effects of the body wall on image quality. Third, the ribs were
artifically placed closer together by 1.7 mm each in the elevation
axis to generate a more challenging intercostal acoustic window
(Fig.~\ref{fig:ribs}, right). These configurations were implemented
simply by modifying the data tagged as bone in the three dimensional
maps described in the companion paper. All other acoustical
properties, including the distribution of fat, connective tissue,
muscle, liver, and sub-resolution scatterer positions, remained
unchanged. The accurate registration of these acoustical maps across
these three imaging scenarios reduces any confounding factors and
simplifies the analysis in terms of a single parameter, i.e. the rib
configuration. An equivalent {\it in vivo} version of this experiment
would not be feasible. If the acoustical maps or properties change
then this would have direct consequences on a number of image quality
parameters such as estimates of aberration, reverberation, speckle
realization, and beamplots.

To generate B-mode images the simulation used exactly the same
propagation physics used by a scanner. A transducer placed at the
surface which emits a focused pulsed wave that propagates through the
heterogeneous acoustic tissue maps. The sound is reflected from the
tissue structures and scatterers. It is subsequently detected at the
transducer surface, and then used to generate ultrasound images.  The
transducer was modeled as a 3.25 $\times$ 1.625 cm 2D array. It
transmitted a 2.5 cycle, 2 MHz, 0.2 MPa pulse focused at 65 mm was
without apodization. A parallel-receive beamforming sequence was
implemented for 5 independent transmit-receive events, as described in
more detail in the companion paper. A 5mm radius circular anechoic
lesion was generated by removing subresolution scatterers in a region
centered at the 65 mm focal depth. This lesion, in conjunction with
measurements of the contrast-to-noise ratio (CNR) provides a basis for
comparing the image quality for the different configurations. The
images were generated with identical transmit-receive sequences so
that any changes in CNR are due to changes in the acoustical field.

\section{Results and discussion}

%***Simulation run time and processing requirements***

\begin{figure}[H]
  \centerline{
  \begin{overpic}[width=0.125\linewidth]{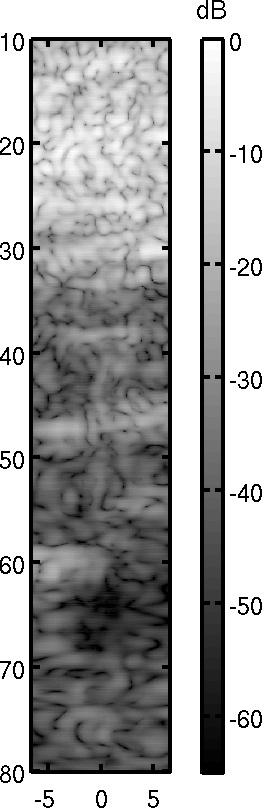}
  \put (16,55) {\textcolor{white}{0.58}}
\end{overpic}
\begin{overpic}[width=0.125\linewidth]{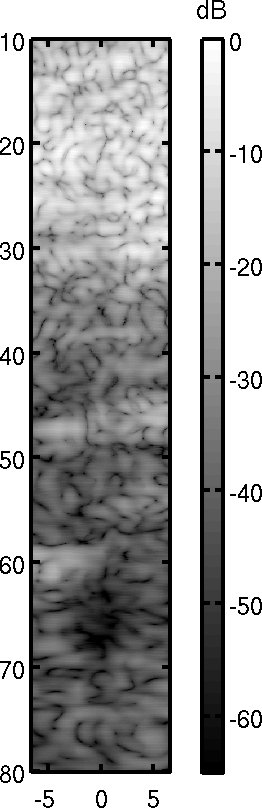}
  \put (16,55) {\textcolor{white}{0.67}}
\end{overpic}
\begin{overpic}[width=0.125\linewidth]{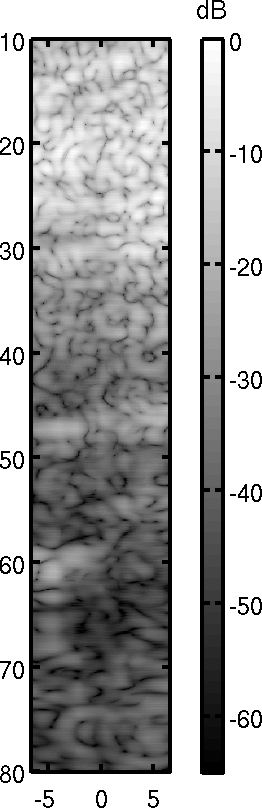}
  \put (16,55) {\textcolor{white}{0.52}}
\end{overpic}
}
%\begin{overpic}[width=\linewidth]{b-mode_composite}
\caption{Fundamental  B-mode images generated by conventional beamforming of the simulated data. The plots show, from left to right, correspond to a scenarios without ribs, with ribs, and with augmented ribs.}
  \label{fig:B-mode}
\end{figure}

\begin{figure}[H]

\centerline{
\begin{overpic}[width=0.125\linewidth]{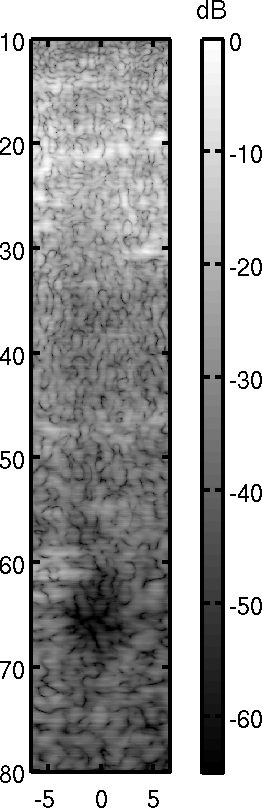}
  \put (16,55) {\textcolor{white}{0.76}}
\end{overpic}
\begin{overpic}[width=0.125\linewidth]{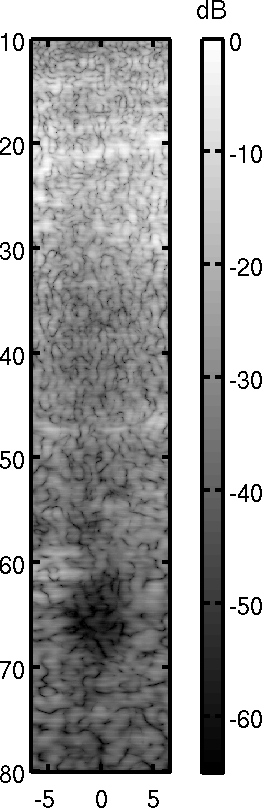}
  \put (16,55) {\textcolor{white}{0.79}}
\end{overpic}
\begin{overpic}[width=0.125\linewidth]{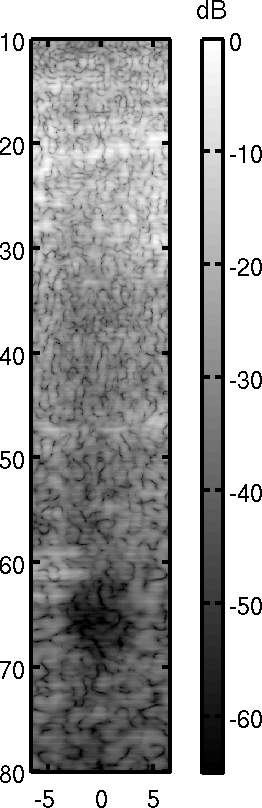}
  \put (16,55) {\textcolor{white}{0.73}}
\end{overpic}
}
%\begin{overpic}[width=\linewidth]{b-mode_composite}
\caption{Harmonic B-mode images generated by conventional beamforming of the simulated data. The plots show, from left to right, correspond to a scenarios without ribs, with ribs, and with augmented ribs.}
  \label{fig:B-mode2}
\end{figure}

The contrast to noise ratio (CNR) of the anechoic lesion was used to
determine the effect of the rib configuration on image quality. The
ultrasound images were generated by applying a conventional
transmit-receive beamforming algorithm to the simulated data received
at the transducer surface.  Fig~\ref{fig:B-mode} shows B-mode images
from left to right, at the fundamental frequency for acoustical maps
without ribs, with ribs, and with augmented ribs.
Fig~\ref{fig:B-mode2} shows this same map configuration with imaging
at the second harmonic frequency. As expected the B-mode images at the
harmonic frequency have a higher CNR, overlaid on the images, than at
the fundamental frequency.

The CNR of the lesion does not strictly decrease when the ribs are
added to the acoustical maps. Counterintuitively the CNR {\it
  improves} with the presence of ribs for both the fundamental and
harmonic images. The highest CNR occurs for the middle case, where the
ribs are in their anatomical position: 0.67 for the image at the
fundamental frequency and 0.79 for the image at the harmonic
frequency. One may expect the highest CNR to occur for the case where
there are no ribs yet it is significantly lower: 0.58 for the
fundamental image and 0.76 for the harmonic image.  Furthermore, in
the augmented rib scenario, on the right, the CNR also decreases
compared to the anatomical case--to 0.52 at the fundamental frequency
and to 0.73 at the harmonic frequency.

This behavior can be understood by analyzing the beamplots, nonlinear
propagation, and reverberation clutter which have different effects on
the image quality depending on whether imaging is performed at the
fundamental or second harmonic frequncy.

%The fundamental (Fig.~\ref{fig:B-mode1}) and harmonic images (Fig.~\ref{fig:B-mode1}) for the tissue maps in Fig.~\ref{fig:ribs} exhibit a counter-intuitive trend in CNR (value overlaid on the images).

%Somewhat more intuitively the case where the ribs have been placed closer together yields the worst CNR: 0.52 for the fundamental image and 0.73 for the harmonic image.
%\begin{overpic}[width=0.5\textwidth,grid,tics=10]{pictures/baum}
%
%\end{overpic}
%\end{document}

\subsection{Beam characteristics}

\begin{figure}[H]
\centering
  \includegraphics[width=0.2\linewidth]{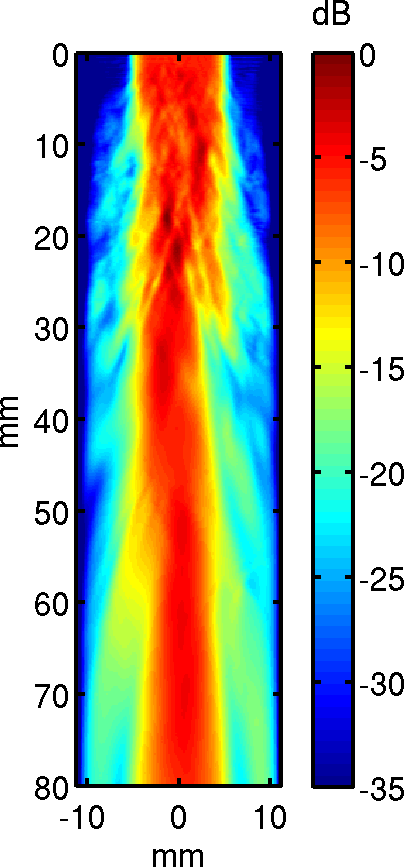}
  \includegraphics[width=0.2\linewidth]{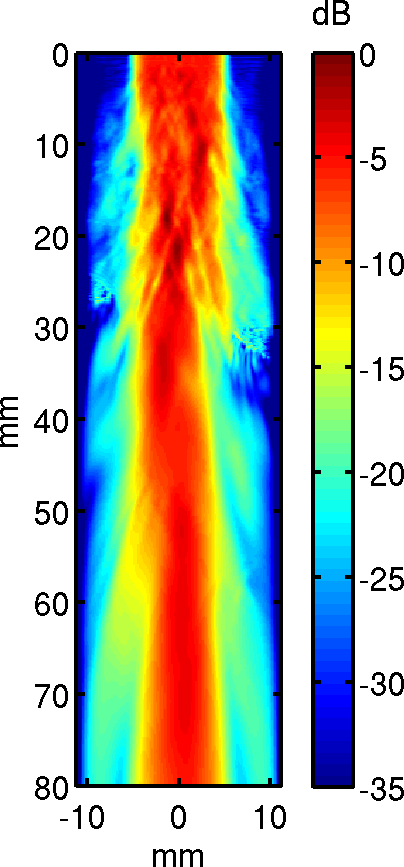}
  \includegraphics[width=0.2\linewidth]{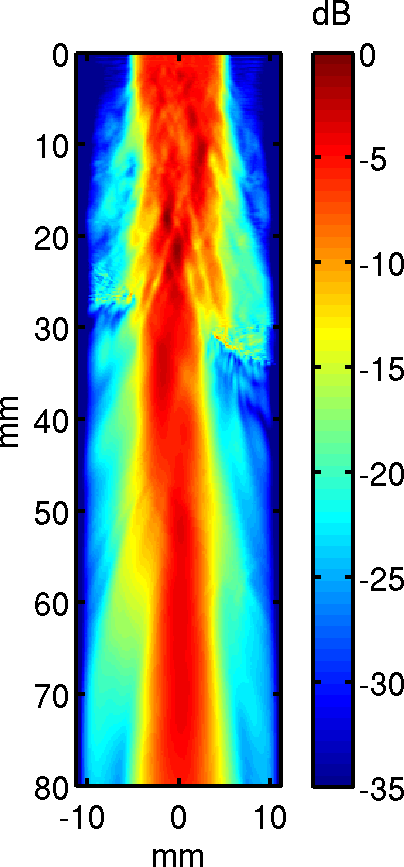}\\

 \caption{The intensity in the elevation plane for the case without
   ribs (left), with ribs (center), and with augmented ribs
   (right) at the fundamental frequency.}
  \label{fig:beamplotel}
\end{figure}

\begin{figure}[H]
\centering
 \includegraphics[width=0.2\linewidth]{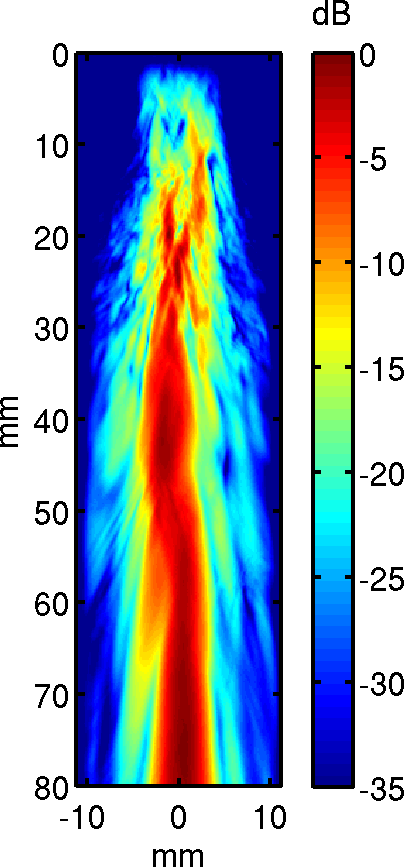}
  \includegraphics[width=0.2\linewidth]{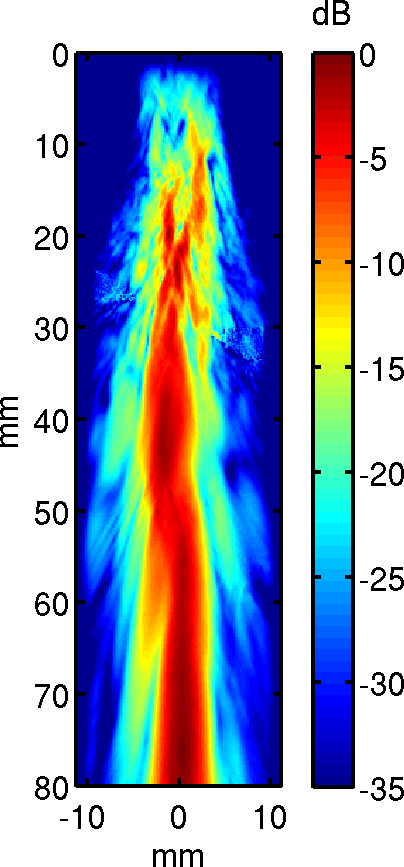}
  \includegraphics[width=0.2\linewidth]{figures/pIel2_liver_new_2MHz_10_moreribs}

 \caption{The intensity in the elevation plane for the case without
   ribs (left), with ribs (center), and with augmented ribs
   (right) at the  second harmonic frequency.}
  \label{fig:beamplotel-b}
\end{figure}

Even though data only from the surface of the human body, at the
transducer location, is used to generate the B-mode images, the
simulation calculates the acoustic propagation throughout the
volume. It is therefore straightforward to determine the imaging beam
intensity by integrating the pressure squared at each location in
space as a function of time. The beam shape, a fundamental property
that predicts image quality in
the central elevation plane for the case without ribs (left), with
ribs (center), and with augmented ribs (right) at the fundamental
(Fig.~\ref{fig:beamplotel}) and second harmonic (Fig.~\ref{fig:beamplotel-b}) frequencies. The lateral beams for
the anatomically correct case is shown in part I of this paper. The
lateral beams for the case without ribs and with augmented ribs are
not shown here because they are not discernibly different.

The middle and right beams in Figs.~\ref{fig:beamplotel},
\ref{fig:beamplotel-b} exihibit a shadowing effect from the ribs at 30
and 35 mm which is apparent from the drop in intensity. The
development of the harmonic energy with propagation can be observed
starting at approximately 10mm of depth.  The harmonic beam is
therefore well developed before it passes in between the ribs,
suggesting that the ribs will also have an impact on the transmitted
harmonic beam. Qualitatively the ribs appears to have a weak influence
on beam shape across the different rib cases, indicating that this is
a good acoustical window.

\begin{figure}[H]
\centering
   \includegraphics[width=0.4\linewidth]{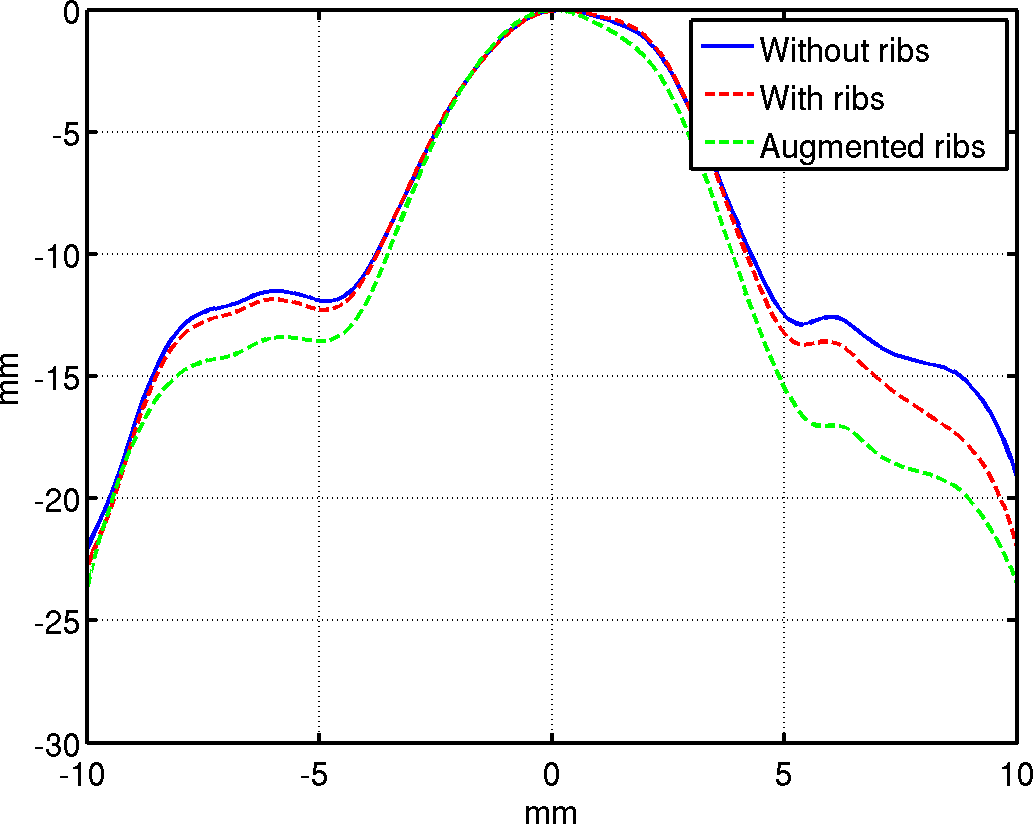}\\
  \includegraphics[width=0.4\linewidth]{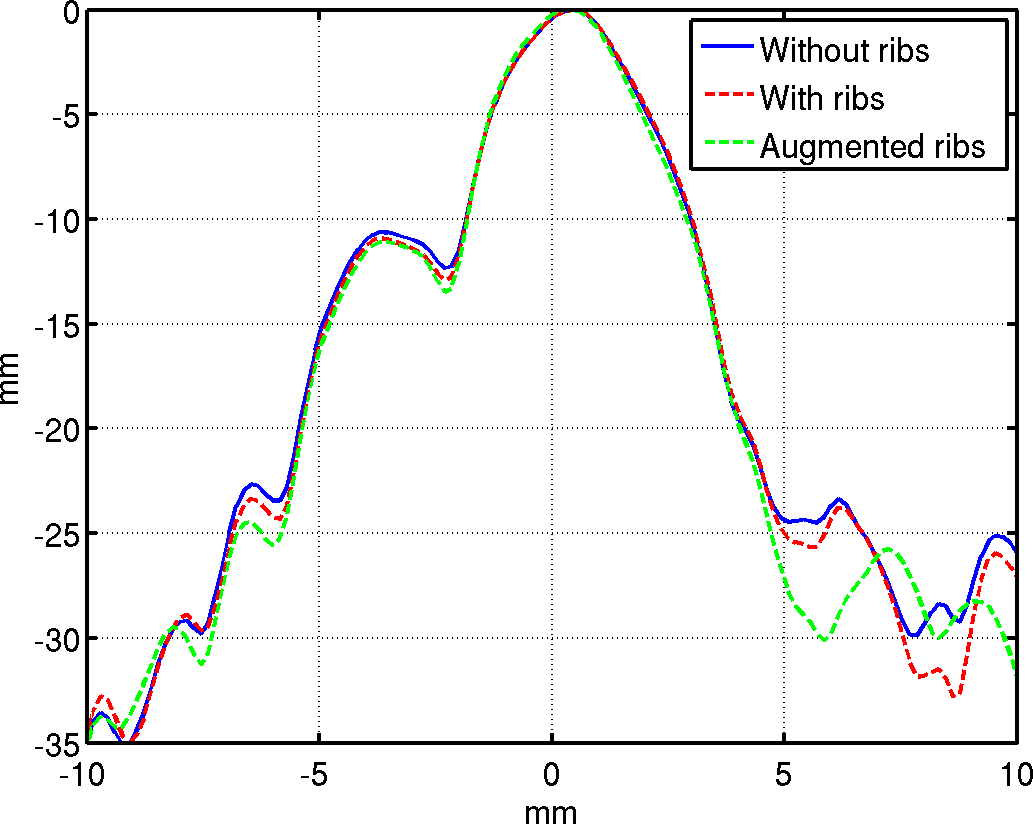}
 \caption{The beamplots at the 65 mm focal depth for the beams in
   Fig.~\ref{fig:beamplotel} for the fundamental (top) and second
   harmonic (bottom) frequencies. The beam characteristics in terms of
   sidelobe level and mainlobe width improve as the influence of the
   ribs increases.}
  \label{fig:beamplotel2}
\end{figure}

A quantitative assessment of the ribs' impact is shown by the
elevation beamplots (Fig.~\ref{fig:beamplotel2}) taken at the focal
depth. Overall the harmonic beams (bottom) have lower sidelobes and a
lower mainlobe compared to the beams at the fundamental frequency
(top). This is consistent with the observed improvements in CNR for
the harmonic images in Figs.~\ref{fig:B-mode},~\ref{fig:B-mode2}.  At the fundamental
frequency the case with ribs (dashed red curve) has sidelobes that are
1-2 dB lower than the no rib case (solid blue curve) and the augmented
ribs beamplot (dashed green curve) has sidelobes that are 3-5 dB lower
and a narrower mainlobe. The beamplots at the harmonic frequency
follow a similar trend: reduction of sidelobes and slight reduction in
beam width. However this trend is less marked, probably because the
narrower harmonic beam passes more easily in between the ribs.

The ribs effectively apodize the beam while maintaining or improving
the mainlobe width. The beamplots therefore predict an {\it
  improvement} with the ribs and an even greater improvement with the
augmented ribs. This prediction is only partially consistent with
trends shown in Figs.~\ref{fig:B-mode},~\ref{fig:B-mode2} where the image quality
decreases with the augemented rib case. The beamplots alone are
therefore not sufficient to explain the observed image quality
metrics.

\subsection{Phase aberration}
%Beamplots integrate the pressure as a function of time and therefore do not have any phase information.

The phase aberration, a fundamental parameter for focusing quality,
was determined just after the body wall and ribs, at a depth of 37 mm
and. It therefore includes the effects of propagation through the near
field skin, connective tissue, fat, and muscle. The phase aberration
was then measured as the deviation of the beam from it's ideal
spherically focused profile, using an intensity thresholded phase
aberration methods described in the companion paper.

% Table
\begin{table}[H]
\centering
\caption{Phase aberration at 37 mm depth.}
\label{tab:a}
\tabcolsep7pt\begin{tabular}{lccc}
\hline
  & {Without ribs}  & {With ribs}  & {Augmented ribs}  \\
\hline
Fundamental & 23.4 ns & 23.4 ns &  23.5 ns   \\
Harmonic & 15.1 ns & 16.2 ns & 16.2 ns  \\
\end{tabular}
\end{table}

The root-mean-square phase aberration for the different
configurations, summarized in Table.~\ref{tab:a}. They are consistent
with the 21.3 ns experimental reports of the arrival time across a
mechanically scanned 2-D aperture for ultrasound propagating through a
human chest wall~\cite{hinkelman1997measurements}. Furtheremore they
are than $\lambda/21$ at the fundamental frequency and $\lambda/14$ at
the second harmonic frequncy, indicating that overall aberration has a
low impact on the image quality\cite{Goodman1996}. The harmonic beams
are significantly less aberrated than fundamental beams. Nevertheless
there is practically no difference in the phase aberration with or
without the ribs. For this imaging scenario the differential influence
of phase aberration on image quality can therefore be ignored.

\subsection{Reverberation clutter}

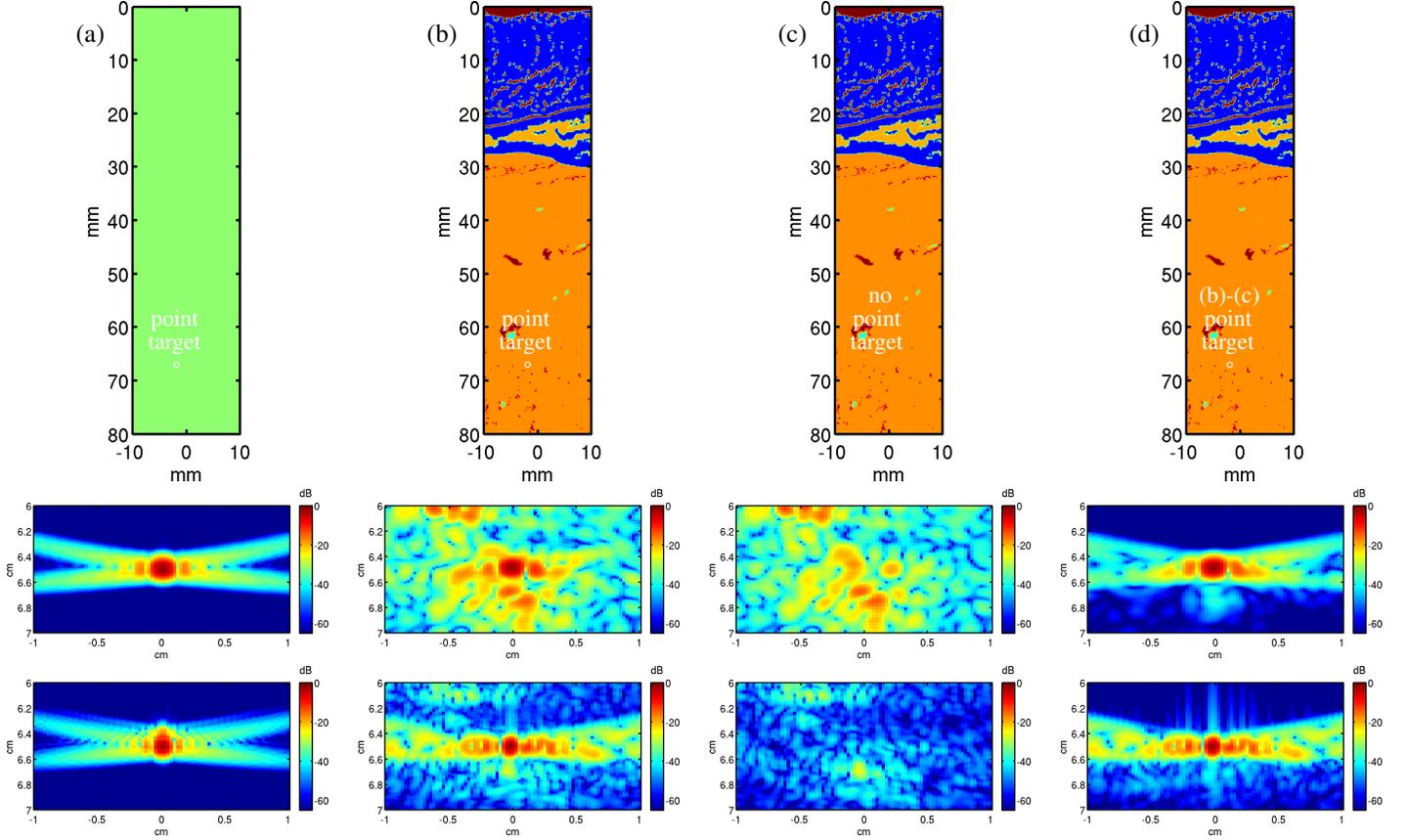
\begin{figure}[H]
\begin{minipage}[b]{\textwidth}
\begin{tabular}{cccc}
\begin{overpic}[width=0.12\textwidth]{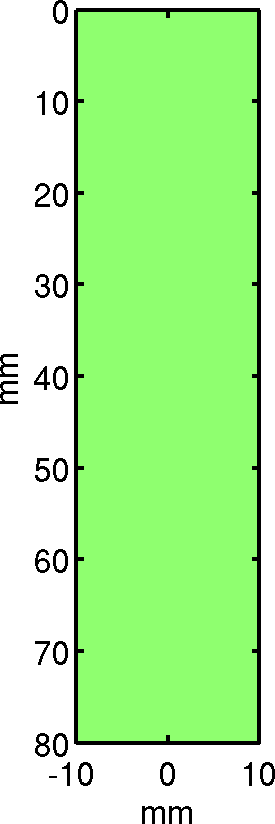}
\put(-5,170){(a)}
     \put(34,45){\bf\textcolor{white}{\circle{2}}}
      \put(24,60){\small\textcolor{white}{ point }}
      \put(23,51){\small\textcolor{white}{ target }}
\end{overpic}&
\begin{overpic}[width=0.12\linewidth]{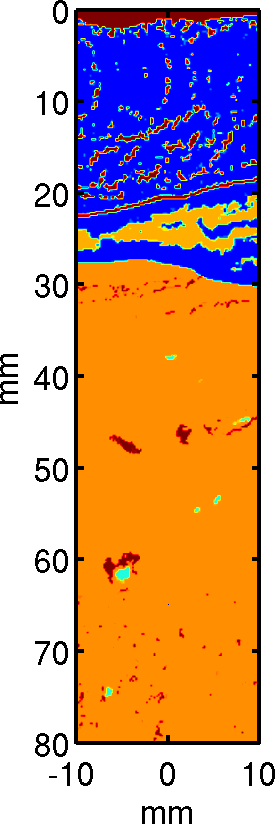}\put(-5,170){(b)}
     \put(34,45){\bf\textcolor{white}{\circle{2}}}
      \put(24,60){\small\textcolor{white}{ point }}
      \put(23,51){\small\textcolor{white}{ target }}
\end{overpic}&
\begin{overpic}[width=0.12\linewidth]{figures/cmap_psf_lat_reduced}\put(-5,170){(c)}
     %\put(34,45){\bf\textcolor{white}{\circle{2}}}
      \put(30,69){\small\textcolor{white}{ no }}
      \put(24,60){\small\textcolor{white}{ point }}
      \put(23,51){\small\textcolor{white}{ target }}

\end{overpic}&
\begin{overpic}[width=0.12\linewidth]{figures/cmap_psf_lat_reduced}\put(-5,170){(d)}
  \put(34,45){\bf\textcolor{white}{\circle{2}}}
     \put(22,69){\small\textcolor{white}{ (b)-(c) }}
      \put(24,60){\small\textcolor{white}{ point }}
      \put(23,51){\small\textcolor{white}{ target }}

\end{overpic}\\
\includegraphics[width=0.24\textwidth]{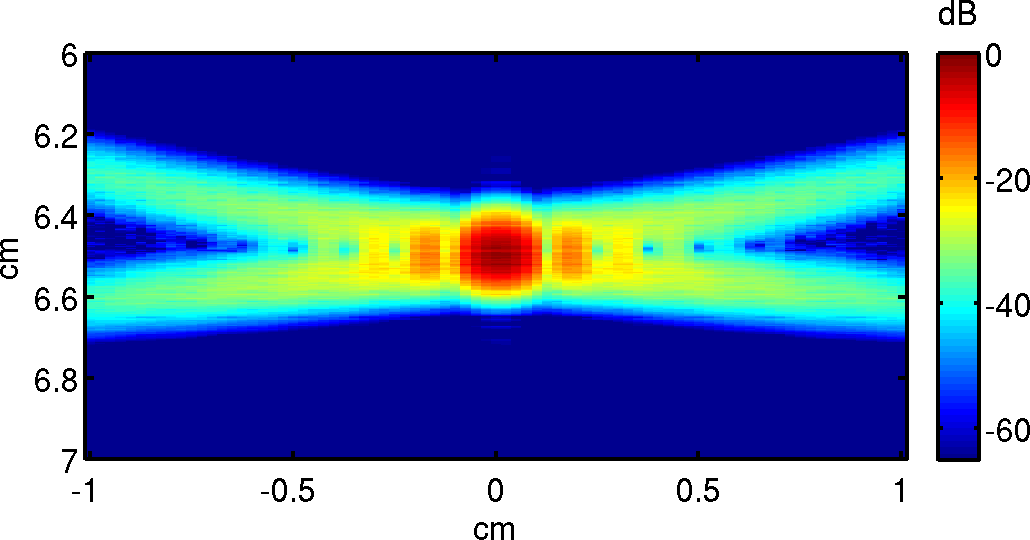}& \includegraphics[width=0.24\textwidth]{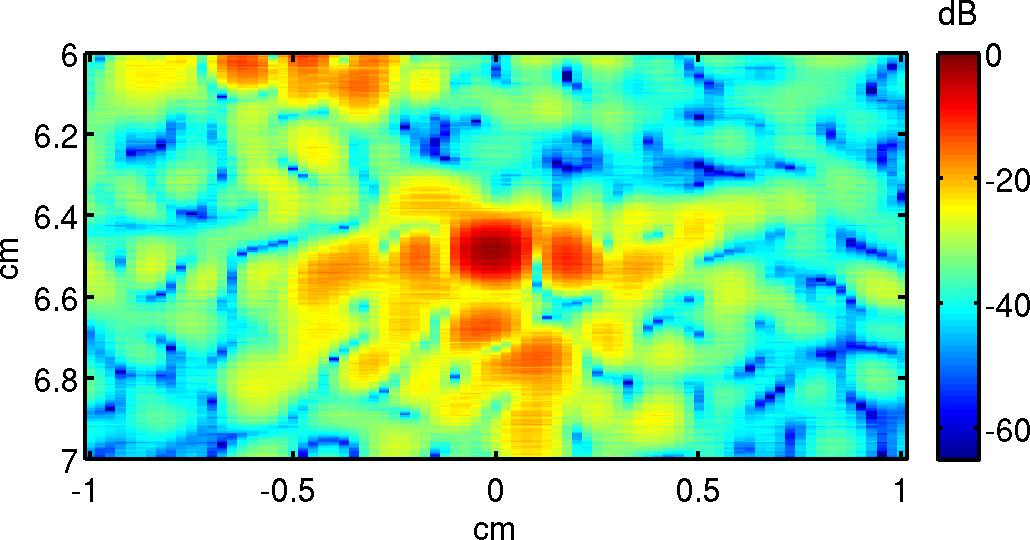} &
\includegraphics[width=0.24\linewidth]{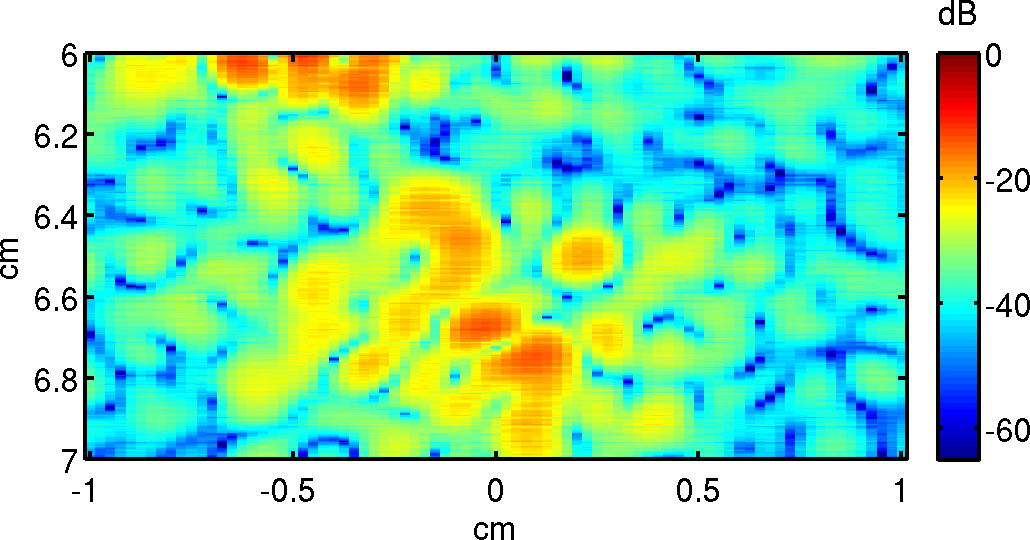} &
\includegraphics[width=0.24\linewidth]{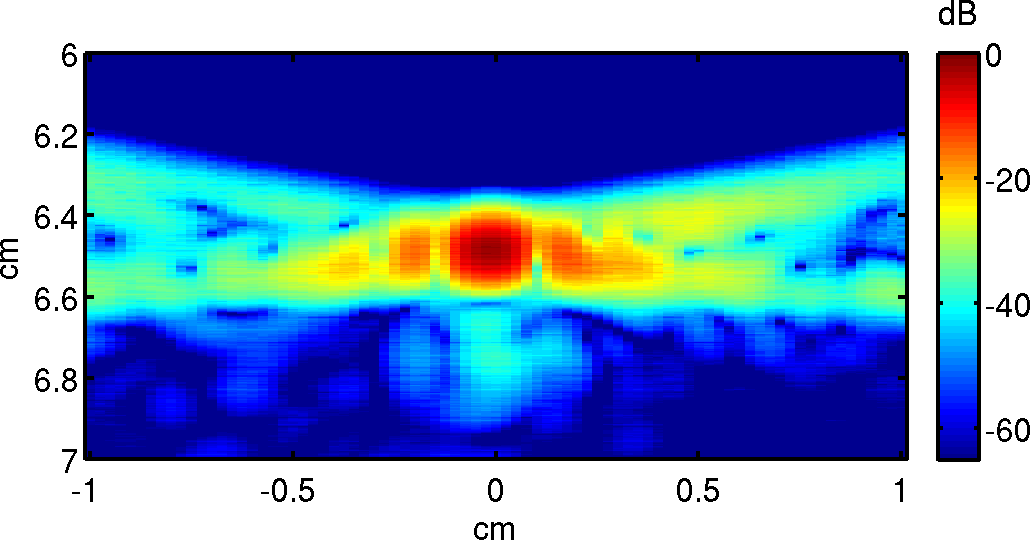}\\

\includegraphics[width=0.24\textwidth]{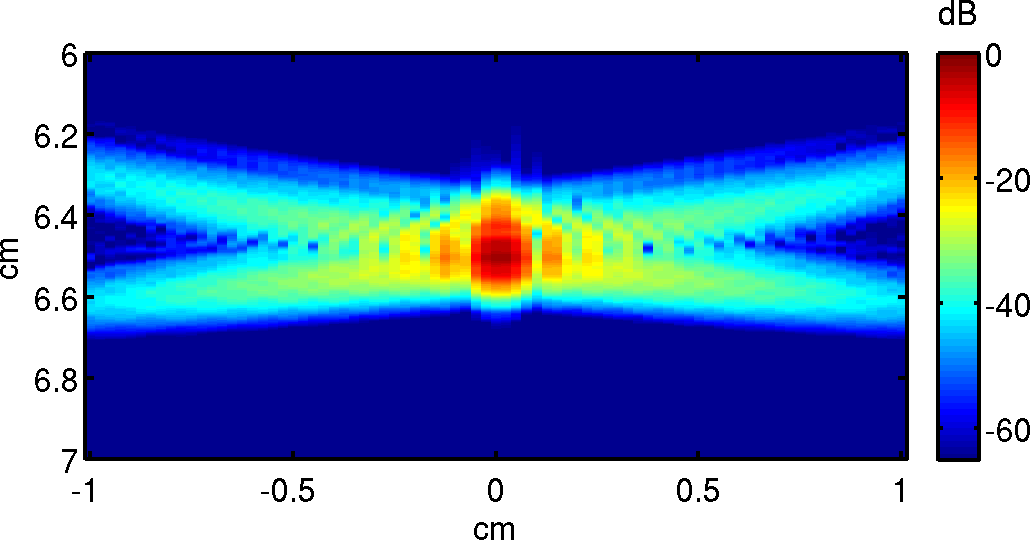}&
\includegraphics[width=0.24\textwidth]{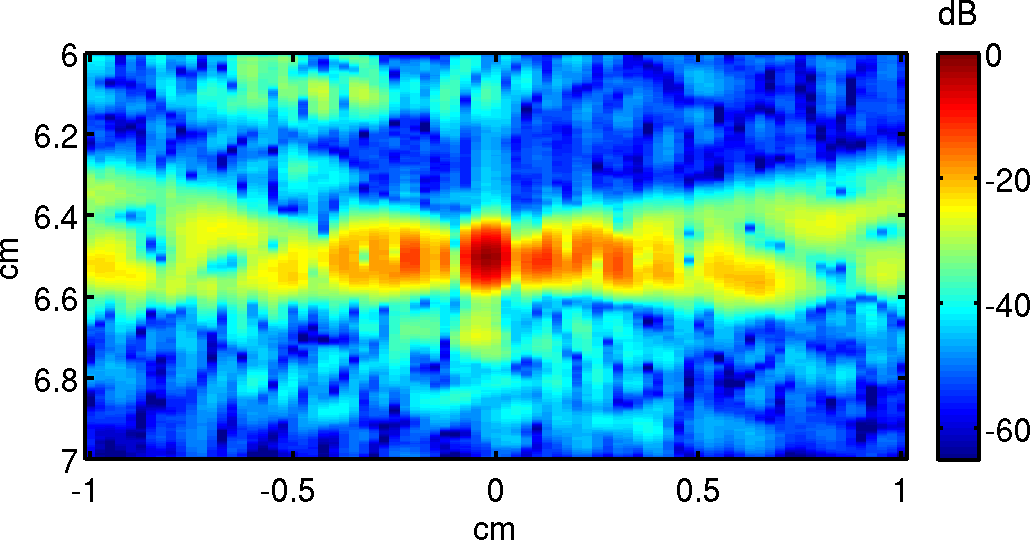}&
\includegraphics[width=0.24\textwidth]{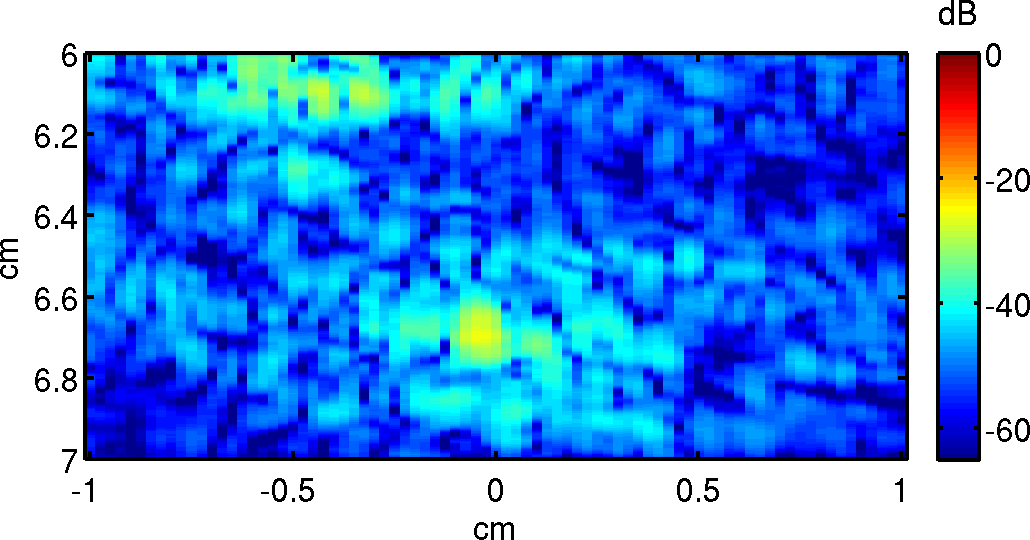}&
\includegraphics[width=0.24\textwidth]{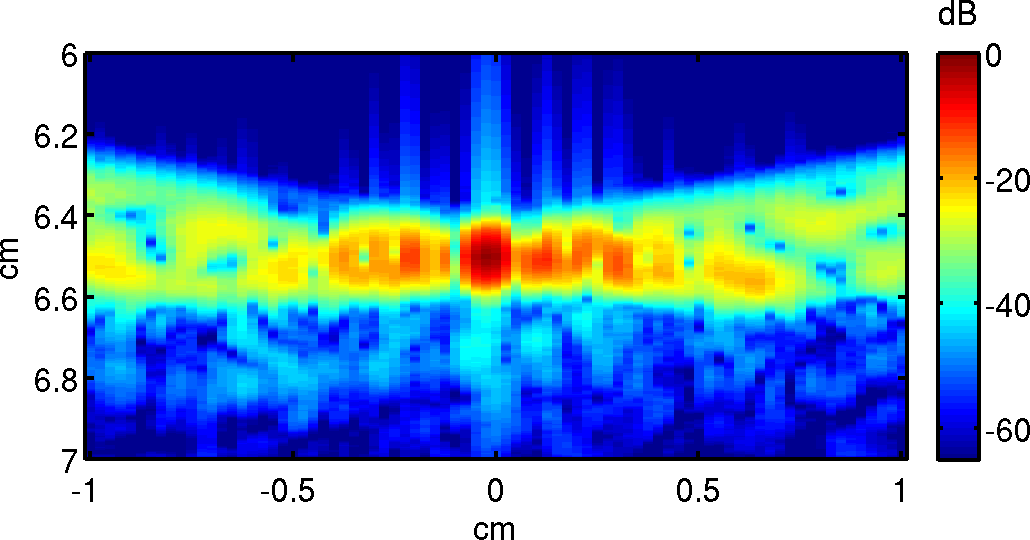}\\
\end{tabular}
\end{minipage}
\caption{Point target simulations for a homogeneous medium (a),
  intercostally (b), intercostally but without a point target at the
  focus (c) and the result of the subtraction of the third column from
  the second which demonstrates reverberation clutter separability
  (d). The top row shows the speed of sound maps. The middle and
  bottom rows are the PSFs for the fundamental and harmonic
  frequencies, respectively.}\label{fig:psf}
\end{figure}

The reverberation clutter was determined with a point spread function
analysis~\cite{pinton2011sources}. A point target was placed at the
focus of a homogeneous material (Fig.~\ref{fig:psf}(a), top) and
single focused transmit-receive simulation was used to generate a
reference point spread functions (PSF) at the fundamental frequency
(Fig.~\ref{fig:psf}(a), middle) and at the harmonic frequency
(Fig.~\ref{fig:psf}(a), bottom). In these images the bow-tie shape of
the isochronous volume is clearly visible. Within the isochronous
volume the signal transmitted by any point on the transducer surface
has had time to travel to the point target and back to any other point
on the transducer surface. In the region above the PSF the signal has
not yet had time to reach the point target and in the region below the
signal it has already passed the point target. Overall, the shapes of
the fundamental and harmonic PSFs are similar. However, the harmonic
PSF is slightly narrower in the lateral dimension and has lower
side-lobes than the fundamental PSF, which is consistent with the
beamwidths shown in the companion paper (Fig. 7) and with the
improvements in image quality observed here (Fig.~\ref{fig:B-mode}).

Then the point target was placed in the human body map
(Fig.~\ref{fig:psf}(b), top) and the same procedure was used to
generate PSFs at the fundamental frequency (Fig.~\ref{fig:psf}(b),
middle) and at the harmonic frequency (Fig.~\ref{fig:psf}(b),
bottom). Note that unlike the B-mode simulations no additional
sub-resolution scatterers were placed in the acoustical maps. The
influence of the body wall is apparent from the speckle pattern that
surrounds the PSF. This speckle pattern does not come from
sub-resolution scatterers near the point target because they were not
included in the acoustical maps. Rather the speckle present in the
region preceding the PSF arises from multipath reverberation because
the signal has not yet had time to travel to the point target and back
on a ballistic path. This suggests that reverberation clutter is
responsible for the bulk of the signal that degrades the PSF. This
hypothesis was tested with the two additional PSF simulations
described subsequently. 

The speckle has a larger amplitude in the fundamental PSF than in the
harmonic PSF, indicating that there is more degradation at the
fundamental frequency.

The third PSF was calculated by removing the point target from the
acoustical maps of the abdomen (Fig.~\ref{fig:psf}(c), top) and by
generating the fundamental (Fig.~\ref{fig:psf}(c), middle) and
harmonic PSFs (Fig.~\ref{fig:psf}(c), bottom) based solely on the
signal from multipath reverberation clutter. Then the fourth PSF
(Fig.~\ref{fig:psf}(d)) was obtained by subtracting the third
reverberation clutter PSF from the second full abdominal PSF. This
fourth PSF does not have reverberation clutter, however it is still
degraded by phase aberration, which appears within the isochronous
volume, and by pulse lengthening, which appears following the
isochronous volume.

Together the third and fourth PSFs illustrate several qualitative
results. Reverberation clutter occurs over a wide spatial extent. The
overall amplitude of the reverberation clutter is significantly larger
at the fundamental frequency than at the harmonic frequency. Pulse
lengthening is more significant at the harmonic frequency than at the
fundamental. Phase aberration within the isochronous is also more
significant than at the fundamental frequency. This is consistent with
results in Table.~\ref{tab:a} where the RMS aberration expressed a
fraction of the wavelength is $<\lambda/21$ at the
fundamental frequency and $<\lambda/14$ at the harmonic frequency.

%here since it
%is the most relevant to the image quality analysis.

%a Four point spread functions were calculated. First, a point target was placed in a homogenous material field. Second using the previously described 3D tissue maps but with the addition of a point target at the focus. Third the focal point target was removed from the 3D tissue maps. By subtracting the third realization from the second realization we were able to demonstrate the separability of the reverberation clutter. The reverberation clutter levels are summarized in Table~\ref{tab:a}, which shows that the ribs add a negligible amount of clutter in the anatomical case, but a significant amount in the augmented case.

\begin{figure}[H]

  \begin{overpic}[width=0.33\linewidth]{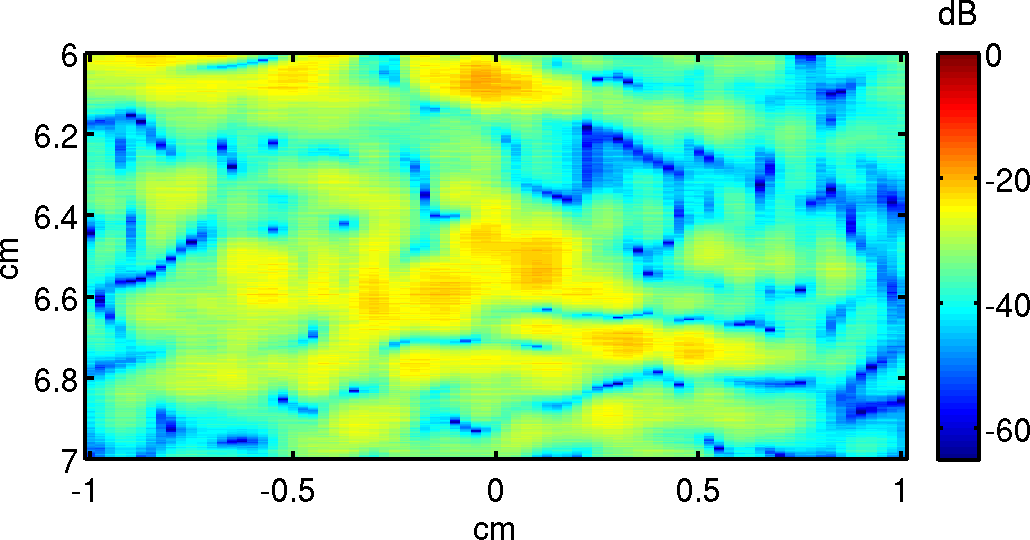}\\
    \put(-5,70){(a)}
  \end{overpic}
    \begin{overpic}[width=0.33\linewidth]{figures/liver_noscat_noribs_10_img1}\\
    \put(-5,70){(b)}
    \end{overpic}
      \begin{overpic}[width=0.33\linewidth]{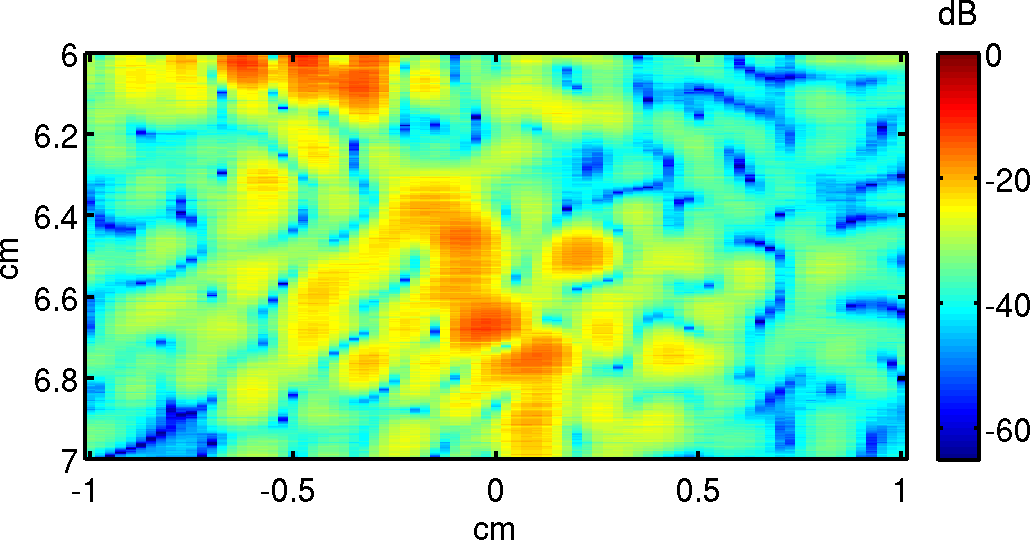}\\
    \put(-5,70){(c)}
  \end{overpic}
\caption{The clutter point spread function at the fundamental frequency for the case without ribs (a), with ribs (b), and with augmented ribs (c).}\label{fig:clutter-psf1}
\end{figure}

\begin{figure}[H]
    \begin{overpic}[width=0.33\linewidth]{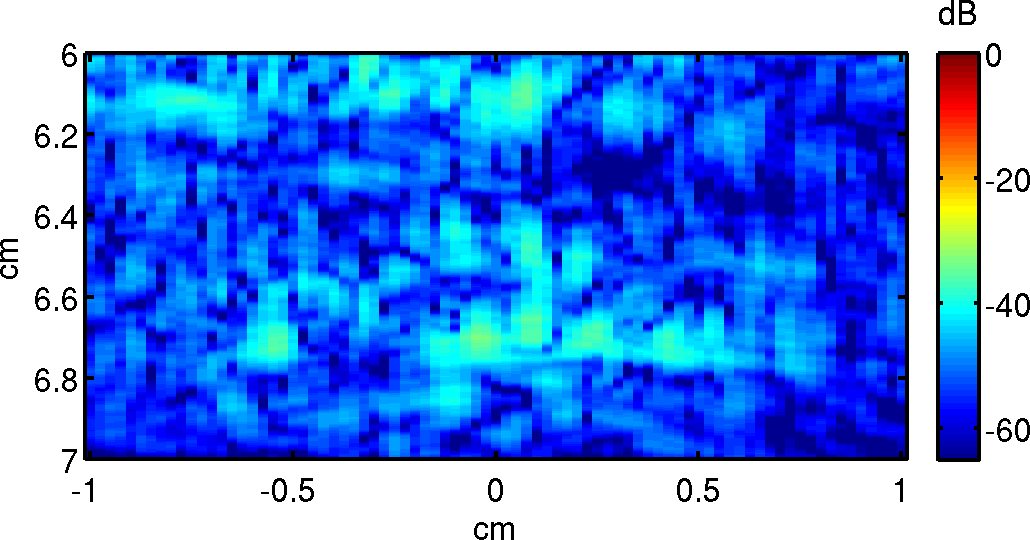}\\
    \put(-5,70){(a)}
  \end{overpic}
    \begin{overpic}[width=0.33\linewidth]{figures/liver_noscat_noribs_10_img2}\\
    \put(-5,70){(b)}
    \end{overpic}
      \begin{overpic}[width=0.33\linewidth]{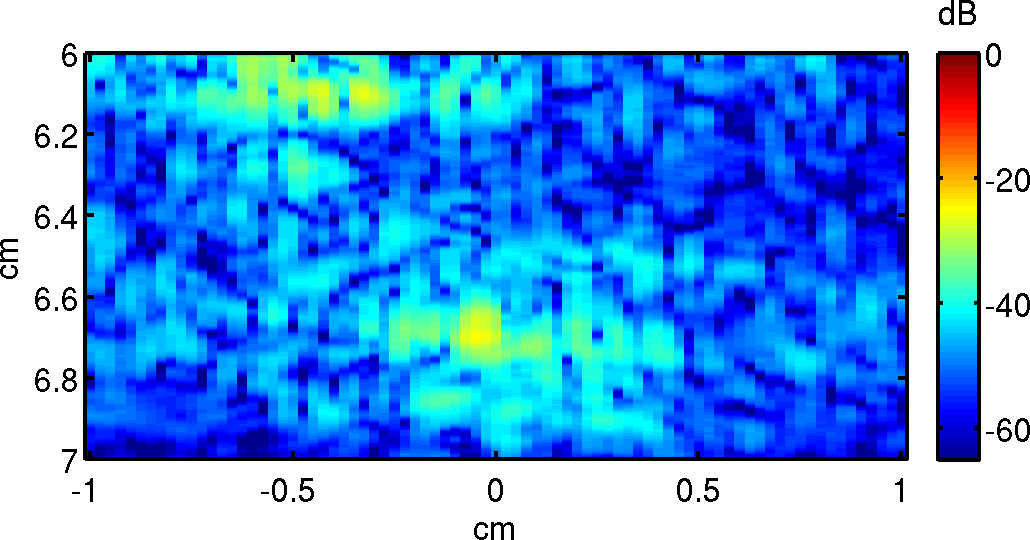}\\
    \put(-5,70){(c)}
  \end{overpic}
\caption{The clutter point spread function at the harmonic frequency for the case without ribs (a), with ribs (b), and with augmented ribs (c).}\label{fig:clutter-psf2}
\end{figure}

%\begin{figure}[h]
%\includegraphics[width=\textwidth]{psf_composite}
%  \caption{laksjd}
%  \label{fig:psfs}
%\end{figure}
This sequence of PSFs was calculated for the three imaging
scenarios. A comparison of the reverberation clutter PSFs for the
fundamental frequency (Fig.~\ref{fig:clutter-psf1}) and harmonic
frequncy (Fig.~\ref{fig:clutter-psf2}) shows how the ribs affect the
overall clutter levels in the ultrasound images. Note that the overall
clutter level is on a dB scale that is calculated relative to the peak
of the abdominal PSF (Fig.~\ref{fig:psf} column b). The average
amplitude of the clutter PSF within the figure for the fundamental
frequency without ribs (Fig.~\ref{fig:clutter-psf1}a) is -35.6~dB. The
addition of ribs (Fig.~\ref{fig:clutter-psf1}b) only slightly
increases the clutter levels to -35.4~dB. However in the augmented rib
case (Fig.~\ref{fig:clutter-psf1}c) the clutter levels are -32.7~dB,
i.e. 2.9~dB higher.

The harmonic clutter PSFs show a similar trend. The average amplitude
of the clutter PSF within the figure for the harmonic frequency
without ribs (Fig.~\ref{fig:clutter-psf2}a) is -51.3~dB. The addition
of ribs (Fig.~\ref{fig:clutter-psf1}b) only slightly increases the
harmonic clutter levels to -51.2~dB. However in the augmented rib case
(Fig.~\ref{fig:clutter-psf1}c) the harmonic clutter levels are
-48.9~dB, i.e. 2.4~dB higher. These results are summarized in
Table~\ref{tab:a}. 


%% CHECK THESE FIGURES %%%
% Table
\begin{table}[h]
\centering
\caption{Reverberation clutter calculated from point spread functions.}
\label{tab:a}
\tabcolsep7pt\begin{tabular}{lccc}
\hline
  & {No ribs}  & {Ribs}  & {Augmented ribs}  \\
\hline
Fundamental & -35.6 dB & -35.4 dB & -32.7 dB    \\
Second harmonic & -51.3 dB & -51.2 dB & -48.9 dB  \\
\end{tabular}
\end{table}

%*** Talk about harmonic beam being fully generated by the time it reaches the ribs***

The clutter PSFs along with the beamplots can be used to explain the
observed B-mode image CNR trends in Fig~\ref{fig:B-mode}. The ribs act
as effective apodizers and reverberators. For the case where the ribs
are in their anatomical position there is enough apodization to
improve the image quality and not enough reverberation to degrade
it. The CNR therefore improves for both the harmonic and fundamental
imaging cases. This is in part due to the fact that the harmonic beam
is fully developed by the time it reaches the ribs
(Fig.~\ref{fig:beamplotel2}). When the ribs are placed artificially
closer together the apodization improves further, but the
reverberation clutter becomes significant and the overall CNR
decreases. Thus, although the overall harmonic image quality is better
than the fundamental image quality, the image quality trends dictated
by body wall's effect on beam shape and multiple reverberation hold
for both frequencies.

\section{Summary and conclusion}

The image quality for intercostal ultrasound was investigated in terms
of beam profile, aberration, and reverberation clutter. For the
specific case considered here, which was derived from the three
dimensional Visible Human anatomical data set, it was shown that the
image quality depended principally on the beam profile and multiple
reverberation. Aberration was shown to be quite small, below
$\lambda/21$ with respect to the fundamental frequency. The
investigation of the beamplots demonstrated that the ribs acted as
effective apodizers, which improved the sidelobes. This shows that if
the main beam is aimed correctly the side lobes can be blocked by the
ribs to obtain an acoustic window that {\it improves} the image
quality. This interpretation was supported by the B-mode images
calculated with the transmit-receive Fullwave simulations which showed
an appreciable increase in CNR at both the fundamental and harmonic
frequencies. This illustrates the importance for a sonographer to find
an appropriate acoustic window.

Then, by changing only one variable, i.e. the rib placement, and by
maintaining all other parameters of the body wall composition, such as
fat and connective tissue distribution constant, it was shown how a
different rib placement can degrade the image quality. In particular
the ribs were placed closer together by just a few millimiters which
had the effect of further improving the beamplot. However the CNR
measured with the transmit-receive Fullwave simulations which showed
an appreciable decrease in CNR at both the fundamental and harmonic
frequencies. It was thus clear that the beamplot and aberration alone
was not sufficient to predict image quality in this case. Multiple
reverbration was therefore calculated using a point spread function
analysis that can separate reverberation from multi-path scattering
from other sources of image degradation. It was shown that when the
ribs were artificially placed closer the degradation from multiple
reverberation decreased the image quality, negating any gains from the
improved beam shape. In conclusion this multiparameter analysis of a
specific intercostal imaging scenario demonstrates the interplay of
two conventional image quality metrics, aberration and beam shape, and
a relatively under-appreciated image quality metric, multiple
reverberation. To estimate the relative contribution of these metrics
on the image quality the full three dimensional wave propagation in
the human body had to be modeled.  The applications of this simulation
tool and image quality analysis are not limited to intercostal imaging
and can be applied to other imaging configurations and to other areas
of the body.

\bibliography{general}
\bibliographystyle{IEEEtran}

\end{document}